\documentstyle[aaspp4]{article}

\begin{document}

\def\gsim{\lower0.5ex\hbox{$\; \buildrel > \over \sim \;$}}
\def\lsim{\lower0.5ex\hbox{$\; \buildrel < \over \sim \;$}}
\def\dm{\mbox{$\dot{M}$}} 
\def\rc{\mbox{$R_{\rm c}$}}
\def\rn{\mbox{$R_{\rm m}$}}
\def\rout{\mbox{$R_{\rm out}$}}
\def\ro{\mbox{$R_{0}$}}
\def\os{\mbox{$\Omega_{\rm s}$}}
\def\oseq{\mbox{$\Omega_{\rm eq}$}}
\def\ws{\mbox{$\omega_{\rm s}$}}
\def\wsc{\mbox{$\omega_{\rm c}$}}
\def\tm{\mbox{$\tau_{\dm}$}}
\def\tv{\mbox{$\tau_{\rm visc}$}}
\def\ts{\mbox{$\tau_{\Omega}$}}
\def\peq{\mbox{$P_{\rm eq}$}}
\def\gcm{\mbox{${\rm G\,cm}^3$}}
\def\gs{\mbox{${\rm g\,s}^{-1}$}}
\def\ergs{\mbox{${\rm erg\,s}^{-1}$}}
\def\tzo{\mbox{T\.ZO\ }}
\def\tzos{\mbox{T\.ZOs\ }}
\def\msun{\mbox{$M_{\odot}$}}

\title{
Can the anomalous X-ray pulsars be powered by accretion?
}

\author{
X.-D. Li\altaffilmark{1}
}
%
\altaffiltext{1}{
Department of Astronomy, Nanjing University, Nanjing 210093, China; 
lixd@nju.edu.cn.}

\begin{abstract}

The nature of the $5-12$ s ``anomalous" X-ray pulsars (AXPs) remains a
mystery.  Among the models that have been proposed to explain the properties
of AXPs, the most likely ones are: (1) isolated accreting neutron stars 
evolved from the Thorne-\.{Z}ytkow objects (\tzos) due to complete 
spiral-in during the common envelope (CE) evolution of high-mass X-ray 
binaries (HMXBs), and (2) magnetars, which are neutron stars with ultra-high 
($\sim 10^{14}-10^{15}$ G) surface magnetic fields. We have critically 
examined the predicted change of neutron star's spin in the accretion model, 
and found that it is unable to account for the steady spin-down observed in 
AXPs. A simple analysis also shows that any accretion disk around an 
isolated neutron star has extremely limited lifetime. A more promising 
explanation for such objects is the magnetar model.

\end{abstract}


\section{Introduction}

In recent years there has been growing evidence that there is a class of
pulsating objects, referred to as braking X-ray pulsars (Mereghetti \& 
Stella \cite{ms95}) or anomalous X-ray pulsars (AXPs, van Paradijs, Taam, 
\& van den Heuvel \cite{pth95}), characterized by the following common 
features (see Stella, Israel, \& Mereghetti \cite{sim98} for a recent 
review): (1) similar spin periods in $\sim 5-12$ s range; (2) steady 
spin-down on a timescale of $\sim 10^4-10^5$ yr; (3) relatively low and 
constant X-ray luminosities of $\sim 10^{35}-10^{36}$ \ergs; (4) very soft 
X-ray spectra, typically described by a combination of a blackbody of 
effective temperature $\sim 0.3-0.4$ keV and a power law with a photon index 
$\sim 3-4$; (5) no detected optical counterpart, some possibly associated 
with supernova remnants. These sources originally include 1E 2259+586, 
1E 1048.1$-$5937, 4U 0142+61 and RX J1838.4$-$0301 (see Mereghetti, Belloni, 
\& Nasuti \cite{mbn97} for an alternative interpretation of this system).  
The recently discovered X-ray pulsars 1E 1841$-$045 (Vasisht \& Gotthelf 
\cite{vg97}), RX J0720.4$-$3125 (Haberl et al. \cite{h97}),
1RXS J170849.0$-$400910 (Sugizaki et al. \cite{s97}) and PSR J1844$-$0258 
(Gotthelf \& Vasisht \cite{gv98}; Torii et al.
\cite{t98}) may also belong to the group of AXPs.

Models proposed to explain the properties of AXPs fall into two main 
categories: (1) Neutron stars accreting from a binary companion with very 
low mass (Mereghetti \& Stella \cite{ms95}), or isolated neutron stars 
accreting from circumstellar debris (e.g., Corbet et al. \cite{c95}). 
In the latter case, in particular, van Paradijs et al. (\cite{pth95}) 
proposed that these pulsars are young ($\sim 10^5$ yr) descendants of the 
\tzos (Thorne \& \.Zytkow \cite{tz77}): neutron stars surrounded by the 
remnants of the CE evolution of HMXBs. Based on this picture, Ghosh et al. 
(1997) argued that the accretion flow onto AXPs from the collapsed envelope 
may consist of two components: a spherical component with low-angular 
momentum, giving rise to the black body emission from a considerable fraction 
of the neutron star surface, and a disk component with high-angular momentum 
responsible for the power-law emission and for the long-term spin-down.  
(2) Magnetars (Thompson \& Duncan \cite{td96}; Vasisht \& Gotthelf 
\cite{vg97}) - neutron stars with a very strong ($\sim 10^{14}-10^{15}$ G) 
magnetic field, in which the magnetic field, rather than rotation, provides 
the main source of free energy, and the decaying field powers the X-ray 
emission.  The observed X-ray luminosities and X-ray spectra in AXPs also 
follow naturally from this model (Thompson \& Duncan \cite{td96}).

Since both kinds of models seem to  present a reasonable description of the 
evolution and the energy source of AXPs, in this paper we examine the 
dynamical properties of the accretion models, in view of the observed secular 
spin change in AXPs.  We focus on the isolated, accreting neutron star model,
since optical, dynamical and evolutionary limits seem to argue against
binarity in AXPs (e.g., van Paradijs et al. \cite{pth95}). Our conclusion is 
that accretion models may not be favored for AXPs. 

\section{The accretion models for AXPs}

The accretion flow around an AXP could be quasi-spherical or in the form of a 
disk, depending on the specific angular momentum carried by the accreting 
material in the collapsed common envelope. As argued by Ghosh et al. (1997), 
disk accretion flow in AXPs reaches the stellar surface in field-aligned flow 
onto the two polar caps of the neutron star that occupy a very small fraction 
of the total surface area, while plasma from quasi-spherical accretion flow 
is not completely disrupted in the stellar magnetosphere, and does not become 
fully field-aligned before it reaches the stellar surface, so that the entire 
stellar surface is available for accretion. To explain the blackbody 
component (which contributes up to $50\%$ of the total X-ray luminosity,  see
Stella et al. \cite{sim98}) with an effective radius of order the stellar 
radius, an exterior 
quasi-spherical accretion flow is generally required. Below we discuss two 
kinds of accretion models for AXPs: the spherical accretion model and the 
two-component (disk + spherical) accretion model.

\subsection{The spherical accretion model}

Because of symmetry of accretion and small angular momentum, a spherical 
accretion flow generally exerts a nearly zero net torque on the accreting 
star.  A possible spin-down mechanism for spherically accreting neutron 
stars was suggested by Illarionov \& Kompaneets (\cite{ik90}) -- when
the X-ray luminosity of a neutron star lies between $\sim 10^{34}$ and
$\sim 10^{36}\,\ergs$, roughly in accordance with those of AXPs, an outflowing 
stream of magnetized matter can be formed within a limited solid angle, the 
neutron star loses angular momentum when the outflow forms so deep as to 
capture the magnetic field lines from the rotating magnetosphere.

The outflow formation is, however, connected with the anisotropy and the
intensity of the hard X-ray emission of the neutron star. It requires that
X-rays from the pulsar heat the accreting matter anisotropically through
Compton scattering, so that the heated matter has a lower density than the
surrounding accreting matter, and flows up by the action of the buoyancy
force (Illarionov \& Kompaneets \cite{ik90}; Igumenshchev, Illarionov, \& 
Kompaneets \cite{i93}). This mechanism is most effective for the 
hard X-ray transients (Be/X-ray binaries) due to their especially hard X-ray 
spectra and high X-ray variability, but fails in AXPs because of their soft 
X-ray spectra and almost constant X-ray luminosities.

Another possible spin-down mechanism for AXPs is the conventional 
``propeller" mechanism (Illarionov \& Sunyaev \cite{is75}) - when the 
magnetospheric radius $\rn$ becomes larger than the corotation radius 
$\rc\equiv [GMP^2/(4\pi^2)]^{1/3}$ (where $G$ is the gravitation constant, 
$M$ mass of the neutron star, and $P$ the spin period), mass accretion 
becomes inhibited by the centrifugal barrier, the star expels the material 
once it spins up the material to the local escape velocity at $\sim \rn$. 
Actually,  a strong support for the idea that X-ray emission in AXPs is 
powered by accretion comes from the argument that AXPs are spinning near the 
equilibrium period $\peq\sim 5-12$ s, corresponding to their observed X-ray 
luminosities of $\sim 10^{35} -10^{36} \,\ergs$ and inferred magnetic field 
strengths of $\sim 10^{11}$ G (Mereghetti \& Stella \cite{ms95}; van Paradijs 
et al. \cite{pth95}).  Note that even in the propeller regime, a
significant amount of the quasi-spherically accreting material might still
leak through ``between the field lines" and reach the neutron star surface
(Arons \& Lea \cite{al80}). Evidence for propeller effects has been found in 
the X-ray pulsars GX 1+4 and GRO J1744-28 (Cui \cite{c97}), and in the soft 
X-ray transient source Aql X-1 (Zhang et al. \cite{z98}). 
In these sources the occurrence of the propeller effect is due to variable
mass transfer rate from the companion star. For AXPs, it is difficult to 
imagine why the (almost constant) accretion rates are so finely-tuned to 
put them at the edge of the propeller regime, i.e., $\rn\sim\rc$: if 
$\rn<\rc$, AXPs should behave similarly as other normal wind-fed X-ray 
pulsars (with random walk like spin change); if $\rn>\rc$, AXPs would be 
characterized by weak X-ray pulsations, contradicted with observations.

\subsection{The two-component accretion model}

Spin-down in disk-fed pulsars is naturally expected in the standard 
magnetized accretion disk model (Ghosh \& Lamb \cite{gl79a,gl79b}), in which 
the (dipolar) stellar magnetic field is assumed to penetrate the disk and 
wound up in the toroidal direction, because of the angular velocity 
difference between the accretion disk and the star. The magnetic field lines 
penetrating the disk between the inner radius $\ro$ of the disk and the 
corotation radius $\rc$ spin up the star, while those penetrating the 
accretion disk outside the corotation radius brake the star. The spin 
evolution of the star is therefore the result of a balance between the 
angular momentum carried by the accreting matter from the disk to the star, 
the magnetic spin-up torque from the accretion disk inside the corotation 
radius, and the magnetic spin-down torque from the accretion disk outside the 
corotation radius, i.e., 
\begin{equation}
I\dot{\Omega}_{\rm s}=\dm(GM\ro)^{1/2}n(\ws),
\end{equation}
where $I$ and $\os$ are the moment of inertia and the angular frequency of 
the neutron star, respectively. The dimensionless torque $n(\ws)$ depends on 
the ``fastness parameter" $\ws$, the ratio between $\os$ and the Keplerian 
angular frequency of disk plasma at $\ro$, 
$\ws\equiv\os/\Omega_{\rm k}(\ro)=(\ro/\rc)^{3/2}$: when $0<\ws<\wsc$, $n>0$,
the star spins up; when $\wsc<\ws<1$, $n<0$, the star spins down.
Here $\wsc$ is the critical value of $\ws$ where the torque vanishes (n=0), 
which may lie in the range $0.71-0.95$ (Wang \cite{w95}; Yi \cite{y95};
Li \& Wang \cite{lw96}). This means that the parameter space for spin-down 
in a disk-accreting neutron star ($0.71-0.95<\ws<1$) is rather small compared 
to spin-up.

The standard accretion disk model, however, cannot be adequately applied to 
AXPs. First, for simplicity of investigation, Ghosh \& Lamb assumed that
the stellar spin axis aligns the dipole magnetic moment, while X-ray pulsars 
are obviously oblique rotators. We mention that Wang (\cite{w97}) recently 
evaluated the torque exerted on an oblique rotating star by a magnetized 
accretion disk, and found that when the dipole inclination angle increases, 
the vertical magnetic flux through the disk decreases, and the spin-down 
contribution to the torque weakens, which is unable to offset the spin-up 
torque for inclinations exceeding some limiting value ($\sim 54^\circ-
67^\circ$), even when the fastness parameter approaches unity. This implies 
that {\it disk-fed X-ray pulsars are more likely to be spinning 
up than spinning down}.

Now consider the two-component (disk + spherical) accretion flow onto the 
neutron star.  When the quasi-spherical flow begins to interact with the 
stellar magnetic field, a magnetosphere is formed around the neutron star, 
the spherical inflow passes through a shock at the magnetospheric boundary,
cools, and enters the magnetosphere through Rayleigh-Taylor instability
(Arons \& Lea \cite{al76,al80}).  Since plasma around the magnetosphere is 
highly diamagnetic, the stellar magnetic field lines would be confined 
inside the magnetosphere, unable to stretch out and to penetrate the disk 
outside the magnetosphere to produce a magnetic torque on the star, as 
originally suggested for the pure accretion disk model by Ghosh \& Lamb 
(\cite{gl79a,gl79b}). 
The fact that the blackbody component contributes a considerable fraction
of the total X-ray luminosity in AXPs implies that the accretion rate in 
the spherical flow is roughly comparable to that in the disk flow, or
$\rn\sim \ro$.  Both the magnetospheric radius $\rn$ and the 
inner disk radius $\ro$ should be no larger than the corotation radius $\rc$
for accretion to occur, the star would always experience a spin-up torque
from the most inner region of the disk inside $\rn$, that is, 
{\it in the two-component accretion model, AXPs should be observed to be 
spinning up rather than spinning down}
\footnote{The enigmatic transition between spin-up and spin-down in the
7.66 s low-mass X-ray binary 4U 1626-67 (Chakrabarty et al. \cite{c97}) seems
to provide an exception to the above argument. It appears to be fed by Roche
lobe overflow from a low-mass ($0.02-0.06\,\msun$) degenerate He or CO dwarf
companion (Verbunt, Wijers, \& Burm \cite{v90}), and its spectrum has a large 
quasi-thermal component, plus an extended hard component, although it has
been excluded from the AXP group by various authors. The existing magnetized
disk models are unable to provide an explanation for such torque reversal.
It was recently suggested that the sense of rotation of the disk in
4U 1626-67 has been changed (Nelson et al. \cite{n97}), possibly because of
a warping instability in the disk due to irradiation from the central X-ray
source (van Kerkwijk et al. \cite{vk98}).}.

\subsection{How long can a disk be sustained?}

If we relax our arguments and assume that some unknown mechanisms could
balance the spin-up torque and bring the pulsars to equilibrium spin, then
there exists an alternative explanation for the secular spin-down in AXPs, 
that is, it is caused by the long-term decrease in the mass accretion 
rate (Ghosh et al. \cite{gaw97}).  Since there is no supply to the disk in 
the AXP phase, the disk accretion rate $\dm$ should show a steady, secular 
decrease on the viscous timescale, $\tm\sim\tv\simeq 2\times 10^5 
(\alpha/0.1)^{-1} (\rout/10^{14}\,{\rm cm})^{3/2}$ yr as the disk dissipates. 
Here $\alpha$ is the conventional disk viscosity parameter (Shakura 
\& Sunyaev \cite{ss73}), and $\rout$ is the outer edge of the disk. Because 
the equilibrium period $\peq$ scales with $\dm$ roughly as $\peq\propto 
\dm^{-3/7}$, at equilibrium, a secular, monotonic spin-down is expected on a 
timescale 
\begin{equation}
\frac{P}{\dot{P}} \simeq (\frac{7}{3})\tm \sim 5\times 10^5 
 (\frac{\alpha}{0.1})^{-1}(\frac{\rout}{10^{14}\,{\rm cm}})^{3/2}\,{\rm yr}, 
\end{equation}
roughly in accordance with the observations if $\rout\sim 10^{13}$ cm.
[Note that in Ghosh et al. (\cite{gaw97}, p720) the quantity relating
 $P/\dot{P}$ and $\tm$ should be $7/3$ rather $3/7$.]

To comment on this scenario, we first point out that Ghosh et al. 
(\cite{gaw97}) have used an inappropriate estimate of the viscous timescale 
in the disk.  These authors assumed an outer radius of the disk as large as
$\sim 10^{14}$ cm, comparable to the radius of the \tzo envelope,
but derived equation (2) by adopting the prescriptions for middle region 
(b) of the Shakura \& Sunyaev (\cite{ss73}) disks, which is limited
to $\sim (4\times 10^7-2\times 10^8)$ cm for mass accretion rates relevant
for AXPs $(\sim 10^{15}-10^{16}\,\gs)$. As a result, $\tm$ was not 
estimated self-consistently. 

A prerequisite for spin-down driven by a decrease in $\dm$ is that, 
for any change 
in the accretion rate, the accretion torque could respond promptly so that 
the star's spin is always around the instantaneous equilibrium, i.e., $\tm$ 
should be much longer than $\ts$, the characteristic timescale on which 
the spin frequency $\os$ evolves towards the equilibrium frequency $\oseq$, 
if the mass accretion rate is constant (Henrichs \cite{h83}).  The latter 
can be estimated in the similar way as in Henrichs (\cite{h83}): Assuming 
a simple form of the dimensionless torque, 
\begin{equation}
n(\ws)=\frac{(1-\ws/\wsc)}{(1-\ws)}
\end{equation}
with an intermediate value of $\wsc\simeq 0.7$, we may use a linear 
approximation of $n(\ws)$ around $\ws=\wsc$,
\begin{equation}
n(\ws)\simeq -5\,\ws+3.5.
\end{equation}
For a constant mass accretion rate, $\ws$ is proportional to $\os$, and
we may write
\begin{equation}
\dot{\Omega}_{\rm s}(t)=a\os(t)+b,
\end{equation}
where
$a\simeq -5\dm\ro^2/I$ and $b\simeq 3.5\dm(GM\ro)^{1/2}/I$. 
Solving for $\os(t)$ we obtain:
\begin{equation}
\os(t)=\oseq-(\oseq-\Omega_0)e^{-t/\ts},
\end{equation}
where $\Omega_0$ is the angular frequency at $t=0$, and $\ts=-1/a=\os/b$
or, numerically,
\begin{equation}
\ts\simeq 8.7\times 10^4 \mu_{29}^{-8/7}L_{35}^{-3/7}\,{\rm yr},
\end{equation}
where $\mu_{29}$ is the magnetic moment in units of $10^{29} \gcm$,
and $L_{35}$ is the X-ray luminosity in units of $10^{35} \ergs$. [Throughout
this paper, we adopt typical values of the mass ($1.4 \msun$) and radius 
($10^6$ cm) for a neutron star.] In Table 1 we have listed the observed values 
of $P$ and $L_{35}$ for four AXPs, 1E 2259+586, 4U 0142+61, 1E 1048.1$-$59 
(Stella et al. \cite{sim98} and references therein) and 1E 1841$-$05 
(Vasisht \& Gotthelf \cite{vg97}). 
Also listed are the derived values of $\mu_{29}$, $\ts$ and $\tm$. 
We use the values of $\mu_{29}$, obtained by
assuming that these pulsars are at equilibrium spin, to estimate $\ts$
through equation (7). The evolutionary timescales of the mass accretion rate
$\tm$ are derived from the observed spin-down timescales, 
$\tm=(3/7)(P/\dot{P})$ (For 1E 1841$-$05, the spin-down timescale is not
known, and is assumed to be less than the age of the associated supernova
remnant Kes 73, $\sim 2000$ years, see Vasisht \& Gotthelf \cite{vg97}). 
From Table 1 it is clearly seen that the assumption $\tm\gg \ts$ collapses in 
at least some of the AXPs with shorter spin-down timescale, say, 
1E 1048.1$-$59, and PSR J1841$-$045. (Adopting other forms for the 
dimensionless torque $n(\ws)$ only slightly changes the values of $\ts$; 
see Henrichs 1983.)

To further examine the fate of the disks around AXPs, one needs
a reasonable estimate of the disk size. This can be made in several ways.
(1) In view of stellar evolution, the characteristic radius of the disk 
formed from the CE evolution
is of order the initial binary separation, $\sim 3\times 10^{11}$ 
cm for an HMXB initially consisting of a $1.4\,\msun$ neutron star and a 
$15\,\msun$ companion with an orbital period of 10 days (Podsiadlowski et al. 
\cite{pcr95}). 
If the accretion disk is acquired when the newly formed neutron star
is kicked toward a stellar companion, the characteristic radius of the disk
at formation is $\sim GM/V_{\rm kick}^2\sim (2\times 10^{10} {\rm cm})
(V_{\rm kick}/10^3\,{\rm km\,s}^{-1})^{-2}$, provided that the density 
gradient scale inside the stellar companion is comparable (Thompson \& Duncan 
\cite{td96}).  (2) Limits on the optical and infrared emission from an 
extended disk may severely constrain disk-fed AXP models (Thompson \& Duncan 
\cite{td96}). Recent deep infrared and optical observations by Coe \& 
Pightling (\cite{cp98}) show that, in the case of
1E 2259+586, the limits of $J\ge 20$ and $K\ge 18.5$ imply a limiting
disk size of $\sim 7\times 10^{10}$ cm for a 2500 K blackbody at 4.5 kpc;
the infrared limits for 4U 0142+62 are similar and imply a disk size of
$7\times 10^9-3\times 10^{10}$ cm for a distance of $0.5-2.0$ kpc.
(3) The current size of the disk can also be restricted by the stability 
analysis. The persistent nature of AXPs indicates that the accretion disks 
(if they exist) are stable against the thermal-viscous instability (Frank, 
King, \& Raine \cite{fkr92}), requiring that the temperature 
at the outer edge of the disk should be higher than the hydrogen ionization 
temperature $\sim 6500$ K. Recent calculations by Dubus et al. (\cite{d99})
reveal that the critical accretion rate in the irradiated disk,
below which no steady disk solution can exist, is given by
\begin{equation}
\dm_{\rm cr}\simeq 1.3\times 10^{15} (\frac{\rout}{10^{10} {\rm cm}})^{2.1}
		   (\frac{C}{5\times 10^{-4}})^{-0.5}\,\gs,
\end{equation}
where the quantity $C$ includes the efficiency of X-ray production from 
accretion, the X-ray albedo of the disk, and the disk opening angle, with 
the typical value of $5\times 10^{-4}$. If the X-ray emission from AXPs is
due to accretion, equation (8) implies that currently
the outer radius of the disk in AXPs is $\rout\sim 10^{10}$ cm, comparable
with the estimates in point (2). If the disk had a larger size,
plasma in the disk outside of $\sim 10^{10}$ cm would always be in the cold, 
neutral state (since there is no mass supply, as in binary systems, to 
increase the surface density in the disk),
with an accretion rate much smaller than in the hot, ionized, inner disk,
and finally would be separated away from the disk.

For region (c) in the Shakura \& Sunyaev (\cite{ss73}) disks, the viscous 
timescale is (e.g., Frank, Raine, \& King \cite{fkr92})
\begin{equation}
\tv\sim 0.1 (\frac{\alpha}{0.1})^{-0.8}(\frac{\dm}{10^{15}\,\gs})^{-0.3}
        (\frac{\rout}{10^{10}\,{\rm cm}})^{1.2}\,{\rm yr},
\end{equation}
which is much shorter than the observed spin-down timescale or the 
estimated age of AXPs, unless the viscous parameter $\alpha$ is extremely 
small \footnote{Note that the estimate of $\tv$ for a stable disk is almost 
independent on irradiation, since the inner structure of the disk is not 
essentially altered by the central X-ray source (e.g., Dubus et al. 
\cite{d99} and references therein), especially in the case of AXPs,
in which most of the energy in X-rays, because of their soft nature, 
is absorbed at only the surface of the disk.}.
This suggests that the disk would dissipate soon after its 
formation, and that evolution of mass accretion rate in the disk is too 
rapid to account for the observed spin-down timescale.

We conclude that an accretion disk in AXPs can live with very short lifetime, 
and is unable to spin down the pulsars as observed.

\section{Discussion and conclusion}

By examining the possible spin-down processes in the hypothesized accretion 
flow around AXPs, we are led to the conclusion that accretion may not be 
favored as the energy source in AXPs, no matter whether it has a spherical or 
disk geometry.  In this sense the magnetar model seems to be more promising, 
since the narrow range of the spin periods and the spin-down timescale can be
naturally accounted for under the assumption that the neutron star is
isolated and has spun down by the torque of a relativistic MHD wind, 
approximated as magnetic dipole radiation (Thompson \& Duncan \cite{td96}).  
This is strengthened by the recent measurements of ultra-high magnetic field 
strengths of $\sim 10^{14}-10^{15}$ G in two soft-gamma ray repeaters SGRs 
1806-20 ({Kouveliotou et al. \cite{k98a}) and 1900+14 ({Kouveliotou et al. 
\cite{k98b}), which possess similar spin periods (7.47 s and 5.16 s 
respectively) to the AXPs.  
However, we note that, it still remains open whether 
all of the sources mentioned in \S 1 form a homogeneous group, and can be 
described by a unified model (for example, there are considerable diversities 
in the spectrum of individual source). In the case of 
1E 2259+586, fluctuations in the spin-down rate seem to be consistent with 
the torque noise measured in accreting X-ray pulsars, supporting the 
accretion model (Baykal \& Swank \cite{bs96}), though they could also be 
explained by neutron star glitch and magnetic field evolution (Thompson \& 
Duncan \cite{td96}).  Especially, the spin trend of the newly discovered 
pulsars (RX J0720.4-3125, 1RXS J170849.0-400910 and PSR J1844-0258) is 
unknown. Further X-ray observations, for example, by measuring the long-term 
stability of the X-ray fluxes, secular trends in the pulse periods, are 
required to secure the classification of these pulsars to the growing family 
of AXPs, and to confirm or reject the proposed theoretical models.

\acknowledgements
I am grateful to the referee for helpful comments and suggestions
that have greatly improved the manuscript.
This work was supported by National Natural Science Foundation of 
China.

\clearpage

\begin{table}
  \caption[]{Observed and derived parameters of four AXPs}
  \begin{tabular}{lrcccr}
  \noalign{\smallskip}
  \hline
  Source         & $P$ (s) & $L_{35}$ & $\mu_{29}$ & $\ts$ (yr)        & 
  $\tm$ (yr)\\
  1E 2259+586    &  6.98   &   2.0    &    2.3    & $2.5 \times 10^4$ & 
  $1.3 \times 10^5$\\
  4U 0142+61     &  8.69   &   2.5    &    3.3    & $1.5 \times 10^4$ & 
  $5.2 \times 10^4$\\
  1E 1048.1$-$59 &  6.44   &   0.2    &    0.7    & $2.8 \times 10^5$ & 
  $6.0 \times 10^3$\\
  1E 1841$-$05   & 11.76   &   3.5    &    5.5    & $7.2 \times 10^3$ & 
  $\lsim 1.0 \times 10^3$ \\
  \noalign{\smallskip}
  \hline
  \end{tabular}
\end{table}

\end{document}